\documentclass[12pt,a4paper]{article}
\usepackage[margin=1.75cm]{geometry}

\usepackage{amsfonts}
\usepackage{amsmath}
\usepackage{amssymb}
\usepackage{bm}
\usepackage{dcolumn}
\usepackage{graphicx}
\usepackage{graphics}
\usepackage[utf8]{inputenc}
\usepackage{latexsym}
\usepackage{url}
\usepackage{ccaption}
\usepackage[linktocpage=true]{hyperref}
\usepackage{xspace} 
\usepackage[usenames]{color}
\usepackage{mathrsfs}
\usepackage{epstopdf}
\usepackage{setspace}
\usepackage{titlesec}
\usepackage{pgfgantt}
\usepackage{sidecap}
\usepackage{enumitem}
\usepackage{eurosym}
\usepackage{inputenc}
\usepackage{cite}
\usepackage{titling}
\usepackage{blindtext}
\usepackage[]{caption}

\title{\bf New Horizons for $\Psi$:\\ Extreme-Mass-Ratio Inspirals \\ in Fundamental Fields}

\author{Francisco Duque}
\date{Max Planck Institute for Gravitational Physics (Albert Einstein Institute) \\ Am Mühlenberg 1, D-14476 Potsdam, Germany}

\begin{document}
\begin{titlingpage}

\maketitle
%
\centering
\includegraphics[width=0.65\linewidth]{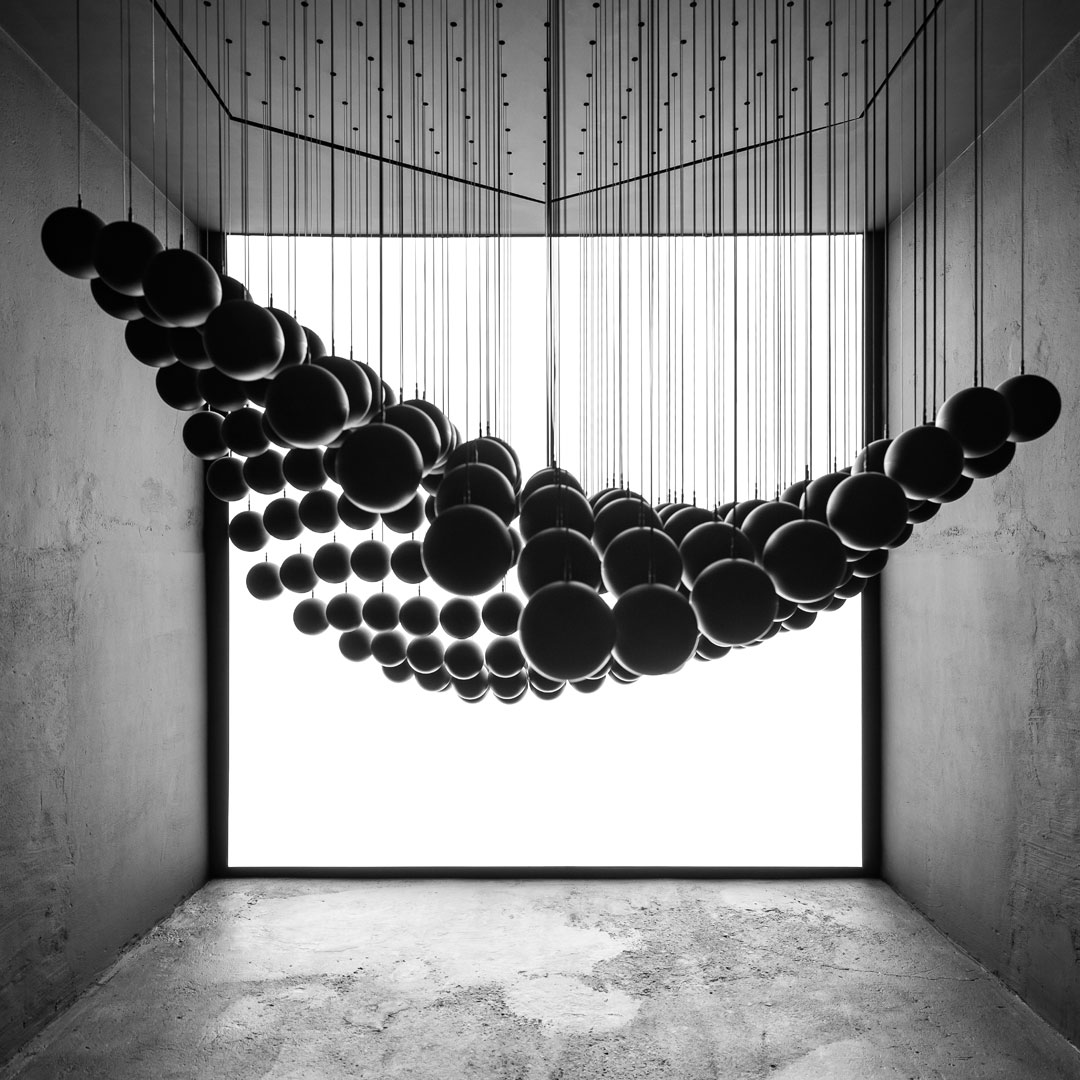}

\textit{INVERSE by Christopher Bauder @ DARK MATTER Berlin (picture by Ralph Larmann)}
%
%
\end{titlingpage}
\tableofcontents
\clearpage
%
\section{Introduction}

This set of notes guided a 2-hour lecture on \textit{Extreme-Mass-Ratio Inspirals (EMRIs) in Fundamental Fields} for the \href{https://strong-gr.com/new-horizons-for-psi/}{New Horizons for $\Psi$ School \& Workshop}, hosted at Instituto Superior Tecnico, University of Lisbon, between 1–5 July 2024. 
Let us start by dissecting the title.

First, \textit{what} are EMRIs? They are binary systems where a stellar-mass \textit{secondary} compact body with mass $m_p$ orbits around a much more massive \textit{primary} object, with mass $M$, such that their mass ratio $\varepsilon = m_p / M \lesssim 10^{-4}$ (systems where $10^{-4} \lesssim  \varepsilon \lesssim 10^{-2}$ are known as \textit{intermediate}-mass-ratios inspirals -- IMRIs)~\cite{Babak:2017tow}.

And \textit{where} can we find such systems? The canonical example is the center of a galaxy, where a population of compact objects like stars, or stellar-mass black holes (BH) is orbiting around a central supermassive BH (SMBH), with $M \gtrsim 10^5 M_\odot$~\cite{Kormendy_2013}. 

And \textit{why} should we care about EMRIs? In 2035, the European Space Agency is going to launch the \textit{Laser Interferometer Space Antenna} (LISA), a constellation of 3 satellites trailing the Earth, separated by $2.5 \times 10^{6}$ km, that will operate as a gravitational-wave (GW) interferometer sensitive to mHz frequencies~\cite{Colpi:2024xhw}. EMRIs complete $\sim10^5$ orbital cycles while in the LISA frequency range and can stay in-band for the entire 4 years of the mission~\cite{Hinderer:2008dm, Babak:2017tow, Barack:2018yly}. This allows measurements of the system parameters, such as SMBH (detector frame) mass and spin, with relative precision as small as $10^{-5}$ (check Tables~$3.6-3.7$ in the LISA Definition Study Report~\cite{Colpi:2024xhw}), an improvement of $10^4$ orders of magnitude compared to current ground-based detectors~\cite{PhysRevX.13.041039}. However, the large number of orbital cycles is also synonymous with extremely complex trajectories and waveforms, which represent one of the hardest challenges in modeling and data analysis for LISA~\cite{Hughes:2021exa, Wardell:2021fyy}. The disparity of mass scales between secondary and primary makes EMRIs almost intractable to tackle with Numerical Relativity (though see~\cite{Wittek:2023nyi, Wittek:2024pis} for recent progress on this). Instead, the \textit{self-force} program solves the EMRI two-body problem by expanding Einstein's equations in the mass-ratio $\varepsilon$~\cite{Pound:2021qin, Barack:2018yvs}. For LISA purposes, this expansion needs to be carried up to $\mathcal{O}(\varepsilon^2)$~\cite{Wardell:2021fyy}.


Moving to the second part of the title, \textit{Fundamental Fields} were the common theme of the school. New ultralight bosonic fields appear naturally in several extensions of GR in a wide range of mass scales $m_{\Psi} \sim 10^{-23}-10^{-1} \, \text{eV}$ and have been proposed as a dark matter (DM) candidate~\cite{PhysRevLett.38.1440, PhysRevLett.40.223, Arvanitaki:2009fg, Hui:2016ltb, Hui:2021tkt, 1985MNRAS.215..575K}. From a technical standpoint, they are the simplest extension of vacuum-GR one can consider, so there is an ``agnostic'', pedagogical aspect in their study. But there is also an astrophysical motivation behind it. When the field's Compton wavelength $ \lambda_\text{c} = 1 / \mu = m_{\Psi} c / h$ is comparable to the BH size, i.e. $ (G M/c^2)  \mu \lesssim 0.5$, the field can extract rotational energy from the BH via \textit{superradiance} (the wave analogous of the Penrose process) and condensate into boson clouds with an hydrogen atom-like structure~\cite{1972Natur.238..211P, Dolan:2007mj, East:2017ovw,East:2018glu, Brito:2015oca,Detweiler:1980uk,Cardoso:2005vk,Arvanitaki:2010sy,Brito:2014wla,Baumann:2019eav}. These clouds are non-spherical and (if the field is real) emit quasi-monochromatic GWs. Ground-based GW detectors, like the LIGO-Virgo-KAGRA collaboration, target stellar-mass BHs, which probe masses around $\mu \sim 10^{-13}\, \text{eV}$~\cite{Brito:2017wnc, Brito:2017zvb, Isi:2018pzk, Yuan:2021ebu, Jones:2023fzz}. On the other hand, the fields can form self-gravitating structures, known as boson stars or DM solitons, which describe well the core of DM halos when they are ``\textit{fuzzy}'', i.e. their mass is $m_{\Psi} \sim 10^{-23}-10^{-19}$ eV~\cite{PhysRev.172.1331, Seidel:1993zk, Lee:1995af, Liebling:2012fv, Torres:2000dw, Schive:2014dra, Schive:2014hza,Veltmaat:2018dfz, Barranco:2017aes, Cardoso:2022nzc}. In this mass range, the de Broglie wavelength is $\gtrsim$ kpc, and thus the field exhibits wave-like behavior, which helps solving some tensions of the canonical $\Lambda$-CDM model at ``small'' length scales, such as the \textit{cuspy-halo} problem~\cite{1994Natur.370..629M, Navarro:1996gj, Weinberg:2013aya}.  

EMRIs are present in galactic centers and we should therefore consider the impact that an ultralight bosonic environment has on their trajectory and subsequent GW signature. Here, we will focus on the simplest system we can think of involving an EMRI and a bosonic environment: a circular inspiral around a non-rotating BH surrounded by a spherical (complex) scalar cloud. This problem has been solved in Refs.~\cite{Brito:2023pyl, Duque:2023cac} (where Ref.~\cite{Duque:2023cac} draws heavily from \cite{Cardoso:2021wlq, Cardoso:2022whc}). Refs.~\cite{Baumann:2021fkf, Baumann:2022pkl, Tomaselli:2023ysb, Tomaselli:2024bdd} also made several important contributions using techniques from Quantum Mechanics, which we will discuss at the end. In addition to the notes, there is a \texttt{Mathematica} notebook with a derivation of the equations of motion and a (rudimentary) ODE solver (\href{https://github.com/FranciscoDuque/EMRIBosons}{https://github.com/FranciscoDuque/EMRIBosons}). My main focus was (trying) to guarantee that who is doing these computations for the first time, in particular students, can follow the logical steps. So, the system we will study is not the most realistic one, and the code I provide is not the most robust (do not be surprised if you start playing with the parameters and something breaks!). But I hope you can understand this simpler case and build from the tools provided if you want to explore further. Occasionally, you will see \textbf{Exercise}s. These are computations that I think someone working in the field should do at least once in their life. 

\pagebreak

\section{Theoretical Setup}

\subsection{Action}

 Our starting point is a minimally coupled \textit{complex} scalar~$\Psi$ evolving on an asymptotically flat spacetime described by the action [we work in natural units $G=c=\hbar = 1$ and use the metric signature $(-+++)$]
\begin{equation}
    S=\int d^4 x \sqrt{-g}\big(\tfrac{1}{16 \pi}R-\tfrac{1}{2}\partial_\mu \Psi\partial^\mu \Psi^*-\tfrac{1}{2}V(\Psi^* \Psi)+ \mathcal{L}^\mathrm{m}\big) , 
\end{equation}
where~$R$ is the Ricci scalar,~$V$ is the self-interaction potential of the scalar field, and $\mathcal{L}^\mathrm{m}$ describes the matter sector, which interacts with the scalar only via the minimal coupling to the metric~$g_{\mu \nu}$.

\textbf{Exercise 1.} Check this theory has a global~$\mathrm{U}(1)$ symmetry with associated Noether current and charge, respectively,
\begin{equation}
    J_Q^\mu=2 g^{\mu\nu} \text{Im} \left[\Psi^* \partial_\nu \Psi\right] \quad, \quad Q= \int_\Sigma d^3x \sqrt{-g}J_Q^t, \label{eq:Noether}
\end{equation}
where $\Sigma$ is a space-like hypersurface, and that varying the action above gives you the Einstein-Klein-Gordon system
\begin{equation}
	G_{\mu \nu}=8\pi (T^\Psi_{\mu \nu}+T^\textrm{m}_{\mu \nu})\qquad , \qquad \Box_g\Psi=\frac{\partial V}{\partial |\Psi|^2}, 
	 \label{eq:EKG}
\end{equation}
where~$\Box_g\equiv \partial_\mu\left[\sqrt{-g}\,\partial^\mu\left(\cdot\right)\right]/\sqrt{-g}$, plus a set of equations for the matter sector. Here,~$G_{\mu \nu}$ is the Einstein tensor and the energy-momentum tensors of the scalar and matter are  
\begin{equation}
T^\Psi_{\mu \nu}= \nabla_{(\mu} \Psi^* \nabla_{\nu)} \Psi-\frac{g_{\mu \nu}}{2}(\nabla^\alpha \Psi^*\nabla_\alpha \Psi+V) \qquad , \qquad T^\textrm{m}_{\mu \nu}=\frac{-2}{\sqrt{-g}} \left[\frac{\delta (\sqrt{-g} \,\mathcal{L}_\textrm{m})}{\delta g^{\mu \nu}}\right] \, . \label{eq:StressTensorScalar}
\end{equation}

As already mentioned, the EMRI's mass asymmetry makes it suitable for a perturbative scheme. We will model the secondary as a point particle of mass $m_p$, whose Lagrangian density and energy-momentum tensor are
\begin{equation}
    \mathcal{L}^\textrm{m}= m_{p} \int d \tau \tfrac{1}{\sqrt{-g}}\,\delta^{(4)}\left[x-x_p(\tau)\right] \quad , \quad 
T_\textrm{m}^{\mu \nu}= m_{p} \int  d\tau \frac{1}{\sqrt{-g}}\delta^{(4)}\left[x-x_p(\tau)\right] \dot{x}_p^\mu \dot{x}_p^\nu \, , \label{eq:StressTensorParticle}
\end{equation}
where $x_p(\tau)$ is the particle worldline parameterized by its proper time~$\tau$. This point particle is sourcing perturbations to a background spacetime which solves the Einstein-Klein-Gordon system at $0$-th order and will represent the primary BH and the scalar field distribution around it. We are interested in finding solutions to first-order in $\varepsilon$
\begin{equation}
    g_{\mu\nu} = \widehat{g}_{\mu\nu} + \varepsilon \, \delta g_{\mu\nu} + \mathcal{O}(\varepsilon^2)\, \quad , \quad 
\Psi = \widehat{\Psi} + \varepsilon \, \delta\Psi + \mathcal{O}(\varepsilon^2) \, , 
\end{equation}
where hatted quantities refer to the background. 

\subsection{The background}

We focus on spherically-symmetric background spacetimes (we had to start somewhere... in the end we will briefly discuss on how to extend this for the rotating case), for which the line element in spherical coordinates and the background scalar field are
\begin{equation}
\text{d}\widehat{s}^2 \equiv \widehat{g}_{\mu \nu} d x^\mu d x^\nu\approx -A(r) dt^2+\frac{dr^2}{B(r)}+r^2  (d\theta^2 + \sin^2 \theta d\varphi^2 ) \quad , \quad \widehat{\Psi} \approx \Psi_0(r) e^{-i \omega t} \, . 
    \label{eq:MetricGenProfile}
\end{equation}
In general, the field's frequency $\omega$ is a complex number obtained from solving the 0-th order Einstein-Klein-Gordon system with appropriate boundary conditions (ingoing at the BH horizon and regularity at large $r$). No-hair theorems prevent the existence of spherically symmetric static solutions, i.e. solutions where $\text{Im}(\omega) \neq0$. However, for sufficiently light fields, the accretion timescale, which is dictated by $\tau_\text{acc} \sim 1/\text{Im}(\omega)$, can be much larger than the observation/inspiral timescale of an EMRI. For DM solitons, whose mass is typically much larger than the BH mass -- our own Milky Way has a supermassive BH with $4.6 \times 10^6\, M_\odot$ and core DM halo estimated to have $\sim 10^9 \, M_\odot$ -- the accretion timescale is~\cite{Cardoso:2022nzc} 
\begin{equation}
    \tau_\text{acc}^\text{soliton} \lesssim \, 10  \left(\frac{10^{10}M_\odot}{M_\Psi} \right)^5 \left(\frac{m_\Psi}{10^{-22}\, \text{eV}} \right)^6  \, \text{Gyr} \, ,
\end{equation}
which can be bigger than the Hubble time ($\tau_\text{Hubble} \sim 14\times 10^{-1} \, \text{Gyr}$). For boson clouds, whose mass is much smaller than the primary, the decay is exponential on timescales (for spherical configurations)~\cite{Cardoso:2022nzc} 
\begin{equation}
    \tau_\text{acc}^\text{cloud} \lesssim  5\times10^9 \left(\frac{10^{8}M_\odot}{M_\Psi} \right)^5 \left(\frac{m_\Psi}{10^{-22}\, \text{eV}} \right)^6 \, \text{Gyr} \, .
\end{equation}

Then, we assume the EMRI is inspiralling, or we are observing it, on a timescale much shorter than this accretion timescale. In this limit, we can take $\text{Im}(\omega) = 0$ (this approximation is also used in both Refs.~\cite{Brito:2023pyl, Duque:2023cac}). For ``true''superradiant clouds, their growth ends in a true bound state around Kerr, with $\text{Re}(\omega) \approx m \Omega_\text{H}$ and $\text{Im}(\omega)=0$, where $\Omega_\text{H}$ is the horizon's angular velocity~\cite{Brito:2014wla}. 

\subsection{The perturbations}

The spherically symmetry of the background spacetime implies that any linear perturbation can be decomposed into irreducible representations of $\text{SO}(3)$. For scalars, this is the standard expansion in spherical harmonics $Y_{\ell m}(\theta, \varphi)$ that one encounters in introductory courses on Quantum Mechanics. We treat $\delta \Psi$ and $\delta \Psi^*$ as independent variables because with this decomposition we obtain equations of motion independent of the coordinate time $t$
\begin{equation}
    \delta \Psi = \frac{1}{r}\sum_{\ell,m} \int d\sigma \, \delta \Psi_+^{\ell m}\left(r\right)  e^{-i (\omega+\sigma) t} Y_{\ell m}(\theta, \varphi) \, \quad , \quad \delta \Psi^* = \frac{1}{r}\sum_{\ell,m} \int d\sigma \, \delta \Psi_-^{\ell m}\left(r\right) e^{-i (\omega-\sigma) t}\ Y_{\ell m}(\theta, \varphi) \, ,  
\end{equation}
where $\sum_{\ell,m}\equiv\sum_{\ell=0}^{\infty}\sum_{m=-\ell}^\ell$ and we move to Fourier domain $t \rightarrow \sigma $. To leave our equations as general as possible before applying them to a particular system, we expand the scalar field potential and its first derivative as 
\begin{equation}
    V \approx \widehat{V} + \sum_{\ell,m} \int d\sigma  \delta V^{\ell m}(r) e^{-i \sigma t} Y_{\ell m}(\theta, \varphi) \quad , \quad U \equiv \frac{\partial V}{\partial |\widehat{\Psi}|^2} \approx  \widehat{U} + \sum_{\ell,m}  \int d\sigma  \delta U^{\ell m}(r) e^{-i \sigma t} Y_{\ell m}(\theta, \varphi) \, .
\end{equation}

Analogously, gravitational perturbations $\delta g_{\mu \nu}$ can be expanded in a basis of ten tensor spherical harmonics (the number of independent components of a rank-2 symmetric tensor). Depending on how they transform under parity transformations $\left(\theta, \varphi \right) \to \left(\pi - \theta, \pi + \varphi \right)$, they are grouped into polar/electric/even (do not change) or axial/magnetic/odd type (pick a $-$ sign)

\begin{align}
    \bm{\delta g}^\text{axial}&=\sum_{\ell,m}\frac{\sqrt{2\ell(\ell+1)}}{r} \bigg[i\,h_{1}^{\ell m }\bm{c}_{\ell m}
-h_{0}^{\ell m} \bm{c}^{0}_{\ell m} + \frac{\sqrt{(\ell+2)(\ell-1)}}{2}h_{2}^{\ell m }\bm{d}_{\ell m} \bigg]\ , \label{eq:DecompositionAxial} \\
\bm{\delta g}^{\text{polar}}&=\sum_{\ell,m} A\,H^{\ell m }_0 \bm{a}^{0}_{\ell m}-i\sqrt{2}H^{\ell m}_1 \bm{a}^{1}_{\ell m} +\frac{1}{B}H_2^{\ell m }\bm{a}_{\ell m} +\sqrt{2}K^{\ell m } \bm{g}_{\ell m} \nonumber \\
&+  \frac{\sqrt{2\ell \left( \ell +1 \right)}}{r}\left(h^{(e) \ell m }_1 \bm{b}^1_{\ell m} - i h^{(e) \ell m }_0 \bm{b}^0_{\ell m} \right) \nonumber \\
& + \left( \sqrt{\frac{(\ell +2 )(\ell +1) \ell (\ell -1 ) }{2} }\bm{f}_{\ell m} - \frac{\ell \left(\ell +1\right)}{\sqrt{2}} \bm{g}_{\ell m} \right) G^{\ell m} \, . \label{eq:DecompositionPolar}
\end{align}
Here, the mode perturbations $h_1^{\ell m}, \, h_0^{\ell m}, \dots$ are functions of only $(t,r)$, while $\bm{a}^0_{\ell m}, \, \bm{a}_{\ell m}, \dots$ are the ten tensor spherical harmonics depending only on $(\theta, \varphi)$, whose explicit form can be found in Ref.~\cite{Sago:2002fe} (or in page 17 of Ref.~\cite{Duque:2023nrf}, and also in the \texttt{Mathematica} notebook).

\textbf{Exercise 2}. Check (at least for some of them) the tensor spherical harmonics are orthonormal on the 2-sphere 
\begin{equation}
\big(\bm{r}^{\ell' m'}, \bm{s}^{\ell m} \big) = \int_{S^2} d\Omega  \big(r^{\ell' m'}_{\mu \nu}\big)^*  s^{\ell m}_{\lambda \rho} \, \eta^{\mu\lambda} \,  \eta^{\nu \rho} \, = \delta_{rs} \delta_{\ell\ell'}\delta_{m m'}\, , 
\end{equation}
where $\eta_{\mu\nu} = \text{diag}\left(-1, 1, r^2, r^2 \sin^2 \theta \right)$. 

The energy-momentum tensor of the particle can also be expanded in this basis
\begin{align}
\bm{T}_p&=\sum_{\ell,m}
{\cal A}^{0}_{\ell m }\bm{a}^{0}_{\ell m}+{\cal A}^{1}_{\ell m}\bm{a}^{1}_{\ell m}
+{\cal A}_{\ell m }\bm{a}_{\ell m}+{\cal B}^{0}_{\ell m }\bm{b}^{0}_{\ell m}
+ {\cal B}_{\ell m }\bm{b}_{\ell m}\nonumber +{\cal Q}^{0}_{\ell m }\bm{c}^{0}_{\ell m}+{\cal Q}_{\ell m }\bm{c}_{\ell m} \nonumber \\ & +{\cal D}_{\ell m }\bm{d}_{\ell m}
+{\cal F}_{\ell m }\bm{f}_{\ell m}+ {\cal G}_{\ell m}\bm{g}_{\ell m} \ .\label{eq:SourceParticleDecomp}
\end{align}
For a given source the expansion coefficients can be obtained by projecting the energy-momentum tensor on the respective spherical harmonic, e.g. ${\cal A}^{0}_{\ell m } = (\bm{a}^{0}_{\ell m}, \, \bm{T})$.

\subsubsection{Point-Particle Source}

\textbf{Exercise 3.} As mentioned, the secondary is treated as a point-particle with energy-momentum tensor given by Eq.~\eqref{eq:StressTensorParticle}. Check that in spherical coordinates it can be rewritten as
\begin{equation}
    T_m^{\mu \nu} = T_p^{\mu \nu} = \int d\sigma \, e^{-i \sigma t}\frac{m_p }{r^2\sin\theta} \sqrt{\frac{B}{A}}\frac{dt_p}{d\tau}\frac{dx^\mu}{dt}\frac{dx^\nu}{dt} \times \delta\left[r-r_p (t) \right]\delta\left[\theta - \theta_p (t)  \right]\delta\left[\varphi - \varphi_p(t) \right]\, . \nonumber 
\end{equation}
At leading order in $\varepsilon$, the secondary follows geodesic motion in the background spacetime determined by $\widehat{g}_{\mu\nu}$. However, part of the orbital energy is dissipated via GWs and the scalar configuration. For small enough $\varepsilon$ (like in EMRIs), the energy dissipated over one orbit is much smaller than the orbital energy, and the secondary evolves \textit{adiabatically} over a succession of geodesics. In other words, the orbital phase evolves on a timescale much shorter than the timescale over which quantities characterizing the orbit change, like its energy and angular momentum. This is the basis of the \textit{two-timescale expansion} in which much of the self-force theory is built~\cite{Hinderer:2008dm, Miller:2020bft}, and this adiabatic \textit{flow} of geodesics corresponds to first-order self-force~\cite{Hughes:2021exa}. Here, we will assume (quasi-)circular motion (again, we had to start with the simplest case) for which
\begin{equation}
r_p(t) = r_p\quad , \quad  \, \theta_p(t) = \frac{\pi}{2}\quad ,\quad  \varphi_p(t) = \Omega_p \, t \quad ,
\end{equation}
where
\begin{equation}
\Omega_p = \sqrt{\frac{A'_p}{2r_p}} \quad , \quad A_p \equiv A(r_p) \, , 
\end{equation}
and the associated energy and angular momentum per unit rest mass are
\begin{equation}
E_p = A_p \frac{dt_p}{d\tau} = \dfrac{A_p}{\sqrt{A_p - r_p^2 \Omega_p^2}} \quad , \quad L_p = r_p^2 \frac{d\varphi_p} {d\tau} = \frac{r_p^2 \Omega_p }{\sqrt{A_p - r_p^2 \Omega_p^2}} \, .
\end{equation}

\textbf{Exercise 4.} Check that (at least for some of them) with the prescription above for the orbital motion, the source-term coefficients appearing in Eq.~\eqref{eq:SourceParticleDecomp} are
\begin{align}
&{\cal A}_{\ell m}={\cal A}_{\ell m}^{(1)}={\cal B}_{\ell m}={\cal Q}_{\ell m}=0 \nonumber \quad , \quad {\cal A}_{\ell m }^{0}=m_p\frac{\sqrt{A B}}{r^2}\, E_p \,Y^\star_{\ell m }\left( \frac{\pi}{2}, 0 \right)\,\delta_r\,\delta_\sigma \quad , \nonumber \\
&{\cal B}_{\ell m }^{0}=i\, m_p \frac{\sqrt{AB}}{r^3\sqrt{(n+1)}}L_p\, \,\partial_\phi Y^\star_{\ell m } \Big|_{(\pi/2, 0)}\, \delta_{r}\, \delta_\sigma\ \quad , \quad {\cal Q}_{\ell m }^{0}=-m_p\frac{\sqrt{AB}}{r^3\sqrt{(n+1)}}L_p\, \,\partial_\theta Y^\star_{\ell m }\Big|_{(\pi/2, 0)} \delta_{r} \, \delta_\sigma \, , \nonumber \\
& {\cal G}_{\ell m }=m_p \frac{\sqrt{AB}}{r^4\sqrt{2}}\frac{L_p^2}{E_p}\, Y^\star_{\ell m}\left( \frac{\pi}{2}, 0 \right) \,\delta_{r}\, \delta_\sigma \quad , \quad {\cal D}_{\ell m }=i \, m_p\frac{\sqrt{AB}}{r^4\sqrt{2n(n+1)}}\frac{L_p^2}{E_p} \, \partial_{\theta\phi}Y^\star_{\ell m} \Big|_{(\pi/2, 0)} \, \delta_{r} \, \delta_{\sigma} \quad , \nonumber \\
&{\cal F}_{\ell m }=m_p\frac{\sqrt{A B}}{r^42\sqrt{2n(n+1)}} \frac{L_p^2}{E_p} \,  \, (\partial_{\phi\phi}-\partial_{\theta\theta}) Y^\star_{\ell m}\Big|_{(\pi/2, 0)} \delta_{r} \, \delta_\sigma \, \quad , 
\end{align}
where 
\begin{equation}
n=\ell(\ell+1)/2-1 \quad , \quad \delta_r = \delta(r-r_p) \quad , \quad \delta_\sigma = \delta(\sigma-m \Omega_p) \, .
\end{equation}
You might immediately notice that for $\ell = 1$, the coefficients ${\cal D}_{\ell m }$ and ${\cal F}_{\ell m }$ appear to diverge... Let us see what is going on.

\subsubsection{Gauge-Invariance}

GR is invariant under diffeomorphisms, which at the linear level correspond to
\begin{equation}
x^\mu \to x'^\mu = x^\mu  + \varepsilon \,\xi^\mu \quad , \quad  \delta g_{\mu\nu} \to \delta g'_{\mu\nu} = \delta g_{\mu\nu} - 2\nabla_{(\mu} \xi_{\nu)} \quad , \quad \delta \Psi \to \delta \Psi'=\delta \Psi- \xi^\mu \partial_\mu \widehat{\Psi}\,,
\end{equation}
where~$\xi^\mu$ is the vector field generating the (infinitesimal) diffeomorphism; we can use its four components to impose some coefficients in the expansion of $\delta g_{\mu\nu}$ to vanish. Note that $\xi^\mu$ can also be expanded in its polar and axial components
\begin{align}
\xi^\mu_\text{polar} &=  \sum_{\ell,m} \big( - \frac{1}{A}  \xi_t^{\ell m}  ,   B \xi_r^{\ell m} ,   0  ,  0\big)Y_{\ell m} + \frac{\xi_\Omega^{\ell m}}{r^2 \sin\theta}  \left(0 , 0 , \sin \theta \,\partial_\theta Y_{\ell m} ,  \partial_\varphi Y_{\ell m} \right) \, , \nonumber\\
\xi^\mu_\text{axial} &= \sum_{\ell,m} \frac{\xi_\text{ax}^{\ell m}}{r \sin \theta} \left(0 , 0 ,  \partial_\varphi Y_{\ell m} , - \sin \theta\, \partial_\theta Y_{\ell m} \right) \, , \nonumber
\end{align}
where the $\xi_t^{\ell m}$, $\xi_r^{\ell m}$, $\xi_\Omega^{\ell m}$, $\xi_\text{ax}^{\ell m}$ are only functions of $(t,r)$. 

\textbf{Exercise 5.} Check that in terms of the tensor spherical harmonics
\begin{align}
2 \bm{\nabla} \bm{\xi} &= \left( 2 \partial_t\xi_t   - A' B \xi_r \right) \bm{a}^{0} - i \sqrt{2} \left( \partial_r \xi_t + \partial_t \xi_r  - \frac{A'}{A} \xi_t \right) \bm{a}^{1} + \left(2 \partial_r \xi_r + \frac{B'}{B} \xi_r \right) \bm{a}\nonumber \\ 
&- i\frac{\sqrt{2\ell\left(\ell+1\right)}}{r}\left(\xi_t + \partial_t \xi_\Omega\right) \bm{b}^{0} + \frac{\sqrt{2\ell\left(\ell+1\right)}}{r}\left(\partial_r \xi_\Omega + \xi_r - \frac{2}{r}\xi_\Omega \right)\bm{b} + \sqrt{2 (\ell+1)\ell} \, \partial_t \xi_\text{ax} \bm{c}^{0} \nonumber \\
&- i \sqrt{2 (\ell+1)\ell}  \left( \partial_r \xi_\text{ax} - \frac{\xi_\text{ax}}{r}\right)\bm{c} + i \frac{\sqrt{2\left(\ell +2\right)\left(\ell +1\right)\left(\ell-1\right)}}{r}\xi_\text{ax} \bm{d} \nonumber \\ &+ \frac{\sqrt{2\left(\ell+2\right)  \left(\ell+1\right) \ell  \left(\ell-1\right) } }{r^2} \xi_\Omega \bm{f} 
+ \frac{\sqrt{2}}{r^2} \left(2rB \xi_r - \ell \left( \ell +1 \right) \xi_\Omega \right)\bm{g}  \, , 
\end{align}
where the prime denotes a derivative with respect to $r$ and, henceforth, we omit the $(\ell, m)$ indices to avoid cluttering. If we pick $\xi^\mu$ judiciously, we can eliminate one component of the metric perturbations in the axial sector and three in the polar sector.

We adopt the \textit{Regge-Wheeler} gauge, which is fixed by setting to zero all terms involving angular derivatives of the highest order~\cite{Regge:1957td, Zerilli:1970wzz} (recall Eqs.~\eqref{eq:DecompositionAxial},~\eqref{eq:DecompositionPolar})
\begin{equation}
    h_2=h_0^{(e)}=h_1^{(e)}=G=0 \, . \label{eq:RWgauge}
\end{equation}
However, for $\ell \leq 1$ this choice does not completely fix the gauge, because some of the tensor spherical harmonics are identically zero: $\bm{b}^0 =\bm{b}=\bm{c}^0=\bm{c}=0$ for~$\ell = 0$, and $\bm{d}=\bm{f}=0$ for $\ell \leq 1$. This additional gauge freedom allows to set one more perturbation function to 0. We choose for
\begin{equation}
\text{$\ell=1$} \quad , \quad K = 0 \, .
\end{equation}
The $\ell \leq 1$ modes do not contribute to the radiative degrees of freedom of the gravitational field, and in vacuum can be removed by a gauge transformation~\cite{Zerilli:1970wzz, Detweiler:2003ci}. However, in the presence of an environment (even if just a point particle), they cannot be completely removed and carry physical meaning through the matter sector. For example, for our bosonic configurations there will be dipolar emission of scalar waves.

\subsubsection{Evolution Equations}

We now have all the necessary machinery to tackle the linearized Einstein equations and obtain  a set of ordinary differential equations (ODEs) for the perturbations. We focus on the polar sector, because it is the one more technically involved and dominates energy emission for (quasi-)circular motion (if you understand the polar sector, I am sure you can work out the axial sector by yourself). Our goal is to manipulate the perturbed Einstein-Klein Gordon system and find a system of 5 coupled ODEs that can be written schematically as 
\begin{equation}
    \frac{d\vec{\mathcal{X}}}{dr} + \hat{\alpha} \vec{\mathcal{X}} = \vec{\mathcal{S}} \, , 
\end{equation}
where $\vec{\mathcal{X}} = \left(H_0,\, H_1, \, K, \, \delta \Psi_+, \, \delta \Psi_- \right)$ and $\vec{\mathcal{S}}$ are the source terms related to the point particle.

\textbf{Exercise 6.} Reproduce the next computations for the $\ell \geq 2$ modes

\begin{itemize}
    \item The $\theta\varphi$-component of Einstein's equations gives an algebraic relation between $H_2$ and $H_0$
    \begin{equation}
    H_2 = H_0 - \frac{8\sqrt{2}\pi\,r^2}{\sqrt{n(1+n)}} \cal F \label{eq:AlgebraicH2H0}
    \end{equation}

    \item The $tr$-component of Einstein's equations gives a $1$st-order ODE for $K$ 
    \begin{align}
    &K' + \left(\frac{2}{r} - \frac{A'}{A} \right)\frac{K}{2} = \frac{H_0}{r} + \frac{i}{\sigma}\left(\frac{n}{r^2} + \frac{B}{r^2} + \frac{B'}{r}  + 4\pi V_0  - \frac{4\pi \omega^2 }{A}\Psi_0^2 + 4\pi B \Psi_0'^2 \right)H_1 \nonumber \\
    & - \frac{4\pi\omega }{\sigma r^2} \left(\Psi_0 + r \Psi_0'\right)\left( \delta \Psi_+ - \delta \Psi_- \right) - \frac{4\pi}{r}  \Psi_0'\left( \delta \Psi_+ + \delta \Psi_- \right) +  \frac{4\pi}{r} \frac{\omega}{\sigma} \Psi_0\left(\delta \Psi'_+ - \delta \Psi'_-  \right) \nonumber \\  &- \frac{8\sqrt{2}\pi\,r}{\sqrt{n(1+n)}} \cal F \, . 
    \end{align}

    \item The $t\theta$-component of Einstein's equations gives a $1$st-order ODE for $H_1$ 
    \begin{align}
    H_1' + \left(\frac{B'}{B} + \frac{A'}{A} \right)\frac{H_1}{2} & = -\frac{i\sigma}{B}H_0 -\frac{i\sigma}{B}K + 8\pi i \frac{\omega }{rB}\Psi_0 \left( \delta \Psi_+ - \delta \Psi_- \right) \nonumber \\ &+  \frac{i\sigma}{B}\frac{8\sqrt{2}\pi\,r^2}{\sqrt{n(1+n)}}\mathcal{F} + \frac{i}{B}\frac{8\pi r}{\sqrt{1+n}} \mathcal{B}_0 \, . 
    \end{align}

    \item The $r\theta$-component of Einstein's equations gives a $1$st-order ODE for $H_0$ 
    \begin{align}
   & H_0' + \left(\frac{A'}{A} -\frac{1}{r}\right)H_0 = -\left(\frac{2}{r} - \frac{A'}{A} \right)\frac{K}{2} - \frac{4\pi\omega}{\sigma r^2} \left(\Psi_0 + r  \Psi_0'\right)\left( \delta \Psi_+ - \delta \Psi_- \right) \nonumber \\ & - \frac{4\pi}{r}  \Psi_0'\left( \delta \Psi_+ + \delta \Psi_- \right) 
   + 4\pi \frac{\omega}{r \sigma}\Psi_0 \left(\delta \Psi'_+ - \delta \Psi'_-  \right) \nonumber \\
   &+ \frac{i}{A r^2\sigma }\left(A(n+B) -r^2\sigma^2 + r A B + 4 \pi r^2 A V_0 - 4\pi r^2 \omega^2 \Psi_0^2 + 4\pi r^2 A B \Psi_0'^2  \right)H_1  \nonumber \\
    &+ \frac{4\sqrt{2}\pi\,r^2}{\sqrt{n(1+n)}}\frac{A'}{A} \cal F \, . 
    \end{align}

We know that in GR the gravitational field has only two degrees of freedom, so the ODEs above cannot be all independent. In fact, the $rr$-component of Einstein's equations yields an algebraic relation between $H_0$, $H_1$ and $K$ which we could use to reduce the 3 ODEs above to only two (2 1st order ODEs $\Rightarrow $ 2 boundary conditions to impose $\Rightarrow$ 2 degrees of freedom). In the \texttt{Mathematica} notebook, I chose to work with the 3 equations above, because it is more direct to input in the inhomogeneous part of the perturbed Klein-Gordon equation
    \begin{align}
    &\delta \Psi_+ '' + \frac{1}{2}\left(\frac{A'}{A} + \frac{B'}{B} \right)\delta \Psi_+ ' \nonumber\\  &+ \frac{1}{2r^2 A B}\left( r\left( 2 r(\sigma + \omega)^2  - A'B\right) - A\left(4+4n + 2 r^2 U_0 + r B' \right) \right)\delta \Psi_+ - \frac{r\Psi_0}{B}\delta U  \nonumber\\
    & = - \frac{r\omega \sigma }{A B}\Psi_0 K + i \frac{r \omega}{B}\Psi_0 H_1' + r \Psi_0' (H_0' - K') \nonumber \\
    & + \frac{i}{2 A^2 B}\left( \omega  \Psi_0 \left( r A B'+ B(4A -r A') \right)  + \Psi_0' 2 r(\sigma + 2 \omega) A B  \right)H_1 \nonumber \\  &+ \frac{1}{2AB}\left( - 2 r \omega \left(\sigma + \omega \right) \Psi_0 + \left( r A B'+ B\left(4A + r A' \right) \right) \Psi_0' + 2 r A B  \Psi_0''\right) H_0 \nonumber \\ &-m_p\frac{4\sqrt{2}\pi r^3}{\sqrt{n(1+n)}}\Psi_0' \mathcal{F}' \nonumber \\ &+m_p\frac{4\sqrt{2}\pi r^2}{\sqrt{n(1+n)}AB} \left( r \sigma \omega \Psi_0 - \left(6 A B + r A' B + r A' B \right)\Psi_0'  - 2 r A B \Psi_0''\right) \mathcal{F} \, .
    \end{align}
and the same equation for $\delta \Psi_-$ but with $\omega \rightarrow  -\omega $. 
\end{itemize}

\vspace{5 mm}

Similar steps can be repeated for the $\ell = 1$ mode, where we also impose $K=0$ to fix the additional gauge freedom (you can find this derivation in the notebook, or try to find it yourself as an \textbf{Exercise}).

\subsubsection{Zerilli master variable}

At this stage, you are probably looking to the equations we have just written and thinking \textit{``uff...this looks ugly...what about all the grandiose poetical clichés on GR's mathematical elegance?''}, especially if you have already had contact with BH perturbation theory and are familiar with the \textit{Zerilli} master function~\cite{Zerilli:1970wzz, Zerilli:1970se, Martel:2005ir}. 

\textbf{Exercise 7.} Check that for $\ell \geq 2$ and in a Schwarzschild background, where the background metric functions are
\begin{equation}
    A(r) = B(r) = 1 -\frac{2M}{r} \quad , \quad  \frac{dr_*}{dr} = \frac{1}{\sqrt{AB}} = \frac{1}{1-2M/r} \, ,  \label{eq:SchwarzschildMetric}
\end{equation}
the set of ODEs governing the gravitational perturbations we wrote above are equivalent to the single master equation for the Zerilli function $Z$
\begin{equation}
    \frac{dZ^2}{dr_*^2} + \left( \sigma^2 - V_Z \right)Z = S_Z \, , 
\end{equation}
with the potential 
\begin{equation}
    V_Z = \frac{2\left(1-2M/r \right)}{r^3}\frac{9M^3 + 3n^2 M r^2 + n^2(1+n)r^3 + 9 M^2 n r}{(3M + \sigma r)}^2 \, ,
\end{equation}
and the relation with the original variables being for example in Eqs.~(17)-(19) of Ref.~\cite{Berti:2009kk}. Here we report only the relation for $K$ because it will prove useful below
\begin{align}
K = \frac{dZ}{dr_*} + \frac{6M^2 + n(1+n)r^2 + 3Mnr}{r^2\left(3M + n r \right)}Z \, . \label{eq:ZerilliK} 
\end{align}

The Zerilli master function is a complex-valued gauge-invariant scalar which controls the radiative degrees of freedom of the gravitational field. For the axial sector, there is an equivalent master wave equation for the \textit{Regge-Wheeler} variable. This is undoubtedly a more elegant solution to the vacuum problem than solving a system of coupled ODEs. Unfortunately, in the presence of matter we do not have a notion of a master variable for the polar sectors. In fact, we adopt the Regge-Wheeler gauge for convenience and historical reasons, but it is likely there is a more suitable combination of variables that makes our system of ODEs simpler (e.g. by exploiting the symmetries of the matter sector also). There is ongoing work on this problem~\cite{Mukkamala:2024dxf}, and hopefully, in the near future, we will have master wave equations for generic (spherically-symmetric) spacetimes.  

\subsubsection{Boundary Conditions}

The only ingredient we are missing to solve our evolution equations is to impose boundary conditions at the horizon and at infinity. For $\ell \geq 2$, all gravitational perturbations, which we denote by $X$, admit wave-like solutions ingoing at the horizon and outgoing at large radius. Since our solver does not extend exactly to these radii, we impose a series expansion of the form~\cite{Macedo:2013jja, Brito:2023pyl}
\begin{equation}
X\left(r\rightarrow \infty \right) = e^{+i\sigma r_*}\sum_{j=0}^{n_\infty} \frac{X^{(j)}_\infty}{r^j}  \quad , \quad X\left(r\rightarrow 2M \right) = e^{-i\sigma r_*}\sum_{j=0}^{n_H} X^{(j)}_H \left(r- 2M\right)^j \, , 
\end{equation}
where the coefficients $X^{(j)}_{\infty,H}$ are obtained by inserting these expansions into the homogeneous equations and solving them in each order in $(r-2M)$ and $1/r$, and set one of the $0$-th coefficient to $1$. In the \texttt{Mathematica} notebook, I only went to the $1$-st order coefficient ($n_\infty = n_H = 1$), but adding more terms will make the results more precise. For $\ell = 1$ it is enough to set all gravitational perturbations to 0 at the horizon. In vacuum it is actually possible to find analytical solutions for them which in the Newtonian limit represent a fictitious force due to the fact that our reference frame centered at the primary BH is a non-inertial one~\cite{Zerilli:1970wzz, Martel:2005ir}. 

For the scalar perturbations, we also impose ingoing waves at the horizon and outgoing or regularity conditions at large distances. Because the field has a (non-zero) mass, the series expansions become~\cite{Macedo:2013jja, Brito:2023pyl}
\begin{equation}
\delta \Psi_\pm \left(r\rightarrow \infty \right) = e^{+ik_\pm r_*} r^{-\nu_+} \sum_{j=0}^{n_\infty} \frac{\delta \Psi_{\pm,\infty}^{(j)}}{r^j}  \quad , \quad \delta \Psi_\pm \left(r\rightarrow 2M \right) = e^{-i\omega_\pm r_*}\sum_{j=0}^{n_H} \delta \Psi_{\pm,H}^{(j)} \left(r- 2M\right)^j \, ,  \label{eq:ScalarBCs}
\end{equation}
where
\begin{equation}
\omega_\pm = \sigma \pm \omega \quad , \quad  k_\pm = \text{sign}(\omega_\pm)\sqrt{\omega_\pm^2 - \mu^2} \quad , \quad \nu_+ = - i M \mu ^2/ k_\pm \, .
\end{equation}
We observe that only for frequencies $\omega_\pm > \mu$ there is emisson to infinity. For smaller frequencies, the energy that is transferred from the binary to the environment is not sufficient to surpass the binding energy of the scalar configuration and perturbations are confined, decaying exponentially at large distances.

\subsubsection{Energy Fluxes and Orbital evolution}

Once we solve the evolution equations we have all the necessary information to compute, up to order $\mathcal{O}(\varepsilon^2)$, the fluxes of gravitational and scalar energy carried away to infinity and absorbed by the primary central BH. Generically, we can write them as 
\begin{equation}
\dot{E}^S_L = -\sigma_L \lim_{r\rightarrow r_L} r^2 \int  d\Omega \, T^S_{\mu\nu}\, \xi^{\mu}_{(t)}\, n^{\nu}_L \, , 
\end{equation}
where 
\begin{equation}
    S = \left\{\Psi, g \right\} \quad , \quad L = \left\{H, \infty \right\}  \quad , \quad \sigma_{\left\{H,\infty\right\}} = -,+ \quad , \quad \xi_{(t)} = \frac{\partial}{\partial t}  \quad , 
\end{equation}
with $S$ being the sector (scalar or gravitational),  $L$ the location (horizon or infinity), $\xi_{(t)}$ the Killing vector field associated to invariance under time translations and $n^{\nu}_L$ a normal vector to the respective hypersurface. For the scalar sector, the energy-momentum is the one in Eq.~\eqref{eq:StressTensorScalar} and for the gravitational one we use Isaacson's effective energy-momentum tensor 
\begin{equation}
T^g_{\mu \nu} = \frac{1}{64\pi}\left< \nabla_\mu \delta g^{\alpha \beta} \nabla_\nu \delta g_{\alpha \beta} \right> \, .
\end{equation}
One can show that in a vacuum background, the (time-averaged) gravitational flux can be written in terms of the Zerilli master function as~\cite{PhysRevD.69.044025}  
\begin{equation}
\dot{E}^g_{\infty, H} = \lim_{r \rightarrow \infty, \, 2M} \frac{1}{64\pi}\sum_{\ell, m} \frac{(\ell+2)!}{(\ell-2)!}\, \sigma^2 \left| Z \right|^2 \, .  
\end{equation} 
Using the relation in Eq.~\eqref{eq:ZerilliK} (and its respective asymptotic limit for $r \rightarrow \infty$ and $r \rightarrow 2M$) we can rewrite these as 
\begin{equation}
\dot{E}^g_{\infty} = \lim_{r \rightarrow \infty} \frac{1}{64\pi} \sum_{\ell, m} \frac{(\ell+2)!}{(\ell-2)!}\, \left| K \right|^2 \quad , \quad \dot{E}^g_{H} = \lim_{r \rightarrow 2M} \frac{1}{64\pi} \sum_{\ell, m} \frac{(\ell+2)!}{(\ell-2)!}\, \left| \frac{4M\,K}{\ell(\ell+1) -i\sigma 4M} \right|^2 \, , 
\end{equation} 

\textbf{Exercise 8.} Using the asymptotic behavior dictated by the boundary conditions in Eqs.~\eqref{eq:ScalarBCs} show that the (time-averaged) scalar fluxes can be written as 
\begin{equation}
\dot{E}^\Psi_{\infty} = \lim_{r \rightarrow \infty} \sum_{\ell, m} \omega_+\, \text{Re}(k_+) \left| \delta \Psi_+ \right|^2 \quad , \quad \dot{E}^\Psi_{H} = \lim_{r \rightarrow 2M} \sum_{\ell, m} \omega_+^2 \left| \delta \Psi_+ \right|^2 \quad . 
\end{equation} 

To compute the orbital backreaction due to this emission of energy, we still need to account for the depletion of the cloud itself. Under our assumptions, $\text{Im}(\omega) = 0$, and no accretion onto the secondary, the mass $M_\Psi$ of the bosonic configuration is related to its Noether charge via (check Appendix of Ref.~\cite{Brito:2023pyl})
\begin{equation}
    M_\Psi = \omega Q \, , 
\end{equation}
and therefore under an adiabatic evolution
\begin{equation}
    \dot{M}_\Psi =  \omega \dot{Q} \, . 
\end{equation}
\textbf{Exercise 9.} Again using the asymptotic behavior of the scalar perturbations, together with the fact the Noether current is conserved, show that  
\begin{equation}
    \dot{Q}_\infty = \lim_{r \rightarrow \infty}  \text{Re}(k_+) \left| \delta \Psi_+ \right|^2 \quad , \quad \dot{Q}_H = \lim_{r \rightarrow 2M} \omega_+ \left| \delta \Psi_+ \right|^2 \, . 
\end{equation}

With this, we can finally write a balance law for the adiabatic evolution of the orbital energy of the secondary 
\begin{equation}
    \frac{dE_p}{dt} = - \dot{E}^g_\infty - \dot{E}^g_H  - (\dot{E}^\Psi_\infty - \omega \dot{Q}_\infty ) - (\dot{E}^\Psi_H - \omega \dot{Q}_H) \, . \label{eq:BalanceLaw}
\end{equation}

Note the additional terms $\omega \dot{Q}$ appearing in the balance-law, which represent the mass loss of the cloud, $\dot{M}_\Psi$, due to the ejection of scalar particles. 

\section{Results}

\subsection{Spherical Bosonic Clouds}
%
\begin{figure}[t]
    \centering \includegraphics[width=0.475\columnwidth]{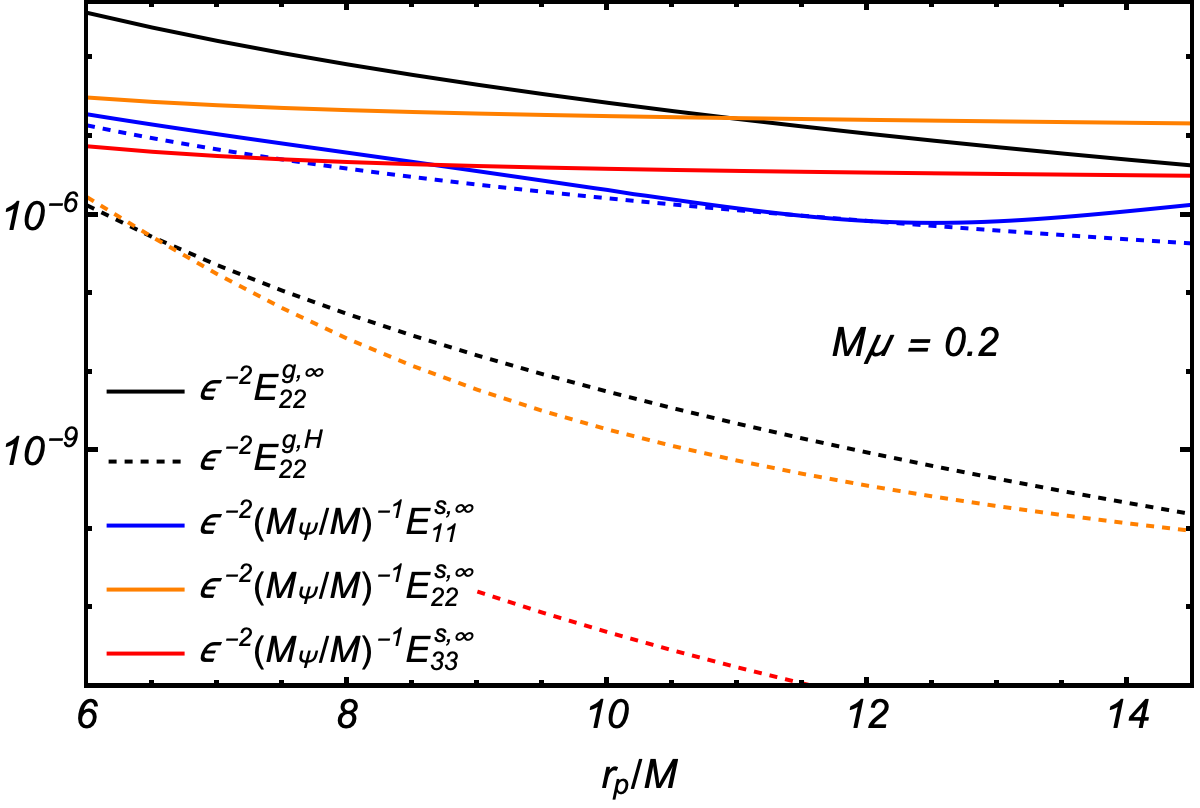}
    \centering \includegraphics[width=0.475\columnwidth]{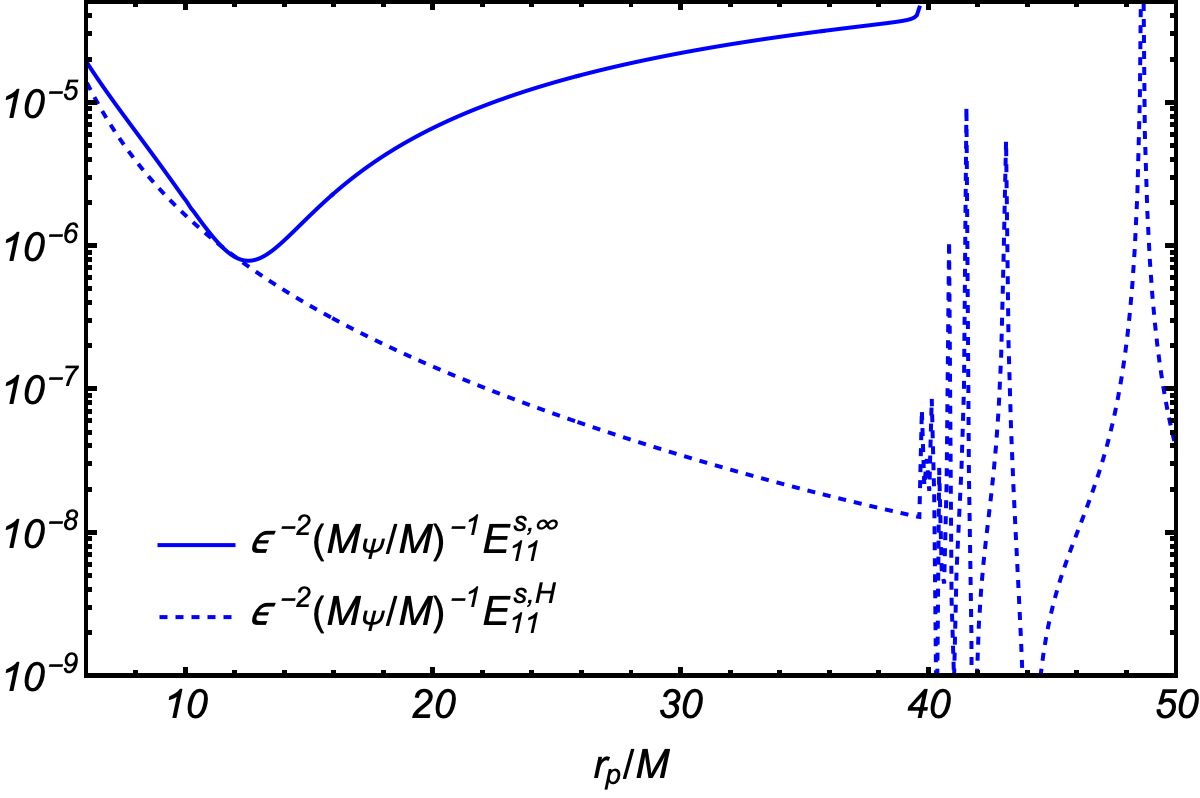}
    \caption{\textit{Left Panel:} Scalar ($s$)/gravitational ($g$) energy fluxes at infinity ($\infty$)/horizon ($H$) as a function of the orbital radius for a circular EMRI immersed in a spherical boson cloud with $M\mu = 0.2$; \textit{Right Panel}: dipolar scalar flux for larger orbital separation.}
    \label{fig:FluxCloud}
\end{figure}

We have all the machinery ready so let us apply it to a particular system. We are going to do it for the simplest system we can think of, which are spherical (ground-state) boson clouds~\cite{Brito:2015oca}. These are (nodless) solutions of the Einstein-Klein-Gordon system in the test-field limit, meaning the backreaction of the (background) field $\hat{\Psi}$ on the (background) geometry is neglected, and the metric reduces to Schwarzschild~\eqref{eq:SchwarzschildMetric}. In this regime, the bosonic cloud has a structure similar to the hydrogen atom with $\alpha = M \mu$ playing the role of a \textit{gravitational fine-structure constant}~\cite{Baumann:2019eav}. For $M \mu \ll 1$, they admit approximate analytical descriptions in terms of Laguerre polynomials.  For the spherical solution~\cite{Brito:2015oca} 
\begin{equation}
    \Psi_0(r) \approx \sqrt{\frac{M_\Psi}{\pi M_{\rm BH}}}(M_{\rm BH} \mu)^2 \left(1-\frac{2M_{\rm BH}}{r}\right) ^{-2i \mu M_{\rm BH}} e^{-M_{\rm BH} \mu^2 r}  \quad , \quad \omega = \mu \left[ 1 - \frac{1}{2}\left(M \mu  \right)^2 \right] \, . \label{eq:BCNewtonianApprox}
\end{equation}
To be fully consistent with the test-field approximation, we neglect all terms related to the scalar field in the gravitational sector of the evolution equations. This is therefore equivalent to solving the vacuum gravitational problem and then using those perturbations as a source of the Klein-Gordon equation. For this, it would have been more practical to work with the Zerilli master function and reconstruct the metric from it. However, with an eye on the generalization for when the test field approximation is not possible, in the \texttt{Mathematica} notebook I still opted to work directly with the set of 5 ODEs. 

I will leave details on the numerical implementation for the \texttt{Mathematica} notebook, and simply jump to the results. In Fig.~\ref{fig:FluxCloud} (left panel), we compare the scalar energy flux carried by different multipoles with the dominant quadrupolar gravitational flux as a function of the orbital radius of the secondary. My solver is not efficient enough, so I was not able to probe very large radii for all multipoles. However, it is clear that at large orbital separations the scalar flux can compete with gravitational radiation. We only show results for one value of $\mu$, but note there is \textit{no} generic analytical scaling of fluxes with $\mu$ that holds up to the relativistic region for small radius (in the Newtonian limit there are some results, check e.g. Eq.~(3.31) of Ref.~\cite{Baumann:2021fkf}). This is more challenging for modeling, but should benefit the detectability of these beyond-vacuum GR effects, since it typically helps breaking degeneracies in the parameter space.   

In the right panel, we show the dipolar fluxes for larger orbital separations. There are obvious ``weird'' features... let us try to understand them.  

\subsection{Gravitational Atoms}

We have discussed that superradiant clouds have a structure similar to the hydrogen atom, which has been explored in depth in Refs~\cite{Baumann:2021fkf, Baumann:2022pkl, Tomaselli:2023ysb, Tomaselli:2024bdd}. We are now going to make use of this analogy to interpret our results.

An energy flux radiated to infinity should correspond to scalar particles escaping the potential well of the bosonic cloud and being ejected from it. In the hydrogen atom, this is simply \textit{ionization}, where the electron is ejected from the atomic nuclei (the fact that in the hydrogen atom we only have 1 particle is what makes the system \textit{quantum}). From the \textit{photoelectric effect}, we know emission is not continuous with respect to the intensity of the external radiation. Instead, the incoming light must have high enough frequency to surpass the binding energy of the electron to the nuclei. Our bosonic cloud follows the same trend. The energy imparted from the binary to the cloud has to exceed a certain threshold, $m\Omega_p > \mu - \omega$, in order to activate the emission of scalar radiation to infinity, \textit{ionizing} the cloud. That is why there is a discontinuity in Fig.~\ref{fig:FluxCloud} for the dipolar flux at $r_p \approx 39.6M$, which corresponds to a frequency $\Omega_p = \sqrt{M/r_p^3} \approx 4.01 \times 10^{-3}$, while $\mu - \omega =  4 \times 10^{-3} $. All modes will have this discontinuity at some radius matching a frequency $(\mu - \omega) / m $, so the larger the radius, the larger the modes contributing to scalar energy emission. 

Moving to the flux at the horizon, we need to understand those peaks. Our cloud is in the \textit{fundamental} mode of a spherically symmetric configuration, i.e. the $\ell = m = 0$ mode with the lowest energy. But there are \textit{overtones} corresponding to excited, higher-energy, spherically-symmetric states. If we were to consider non-spherical clouds, then again in analogy with the hydrogen atom, there will be energy splitting between modes with different angular momentum (labelled by $\ell, \, m$)~\cite{Brito:2015oca}
\begin{equation}
\omega_{n\ell m} = \mu\left(1 - \frac{\alpha^2}{2n^2} - \frac{\alpha^4}{8n^4} - \frac{\left(3n-2\ell_i - 1\right)\alpha^4 }{n^4(\ell_i + 1/2)} + \frac{2(a/M)m_i\alpha^5}{n^3\ell_i(\ell_i + 1)(\ell_i + 1/2)} + \mathcal{O}(\alpha^6)\right) \, , \label{eq:EnergyGravitationalAtom}
\end{equation}
where $n=1,2,...$ labels the overtone and we also included the \textit{hyperfine} splitting ($\Delta m \neq 0 $) when the BH is rotating ($a \neq 0$)~\cite{Tomaselli:2024bdd}. If the binary has orbital frequency matching the energy difference between two states, it induces a resonant transition between them. Those are precisely the peaks we are observing in the right panel of Fig.~\ref{fig:FluxCloud}. The binary transfers energy to the cloud, inducing resonant transitions from the fundamental $n=1,\, \ell = m = 0$ to higher overtones, which decay faster than the fundamental one and are quickly absorbed by the BH. 

Spherical clouds cannot grow via superradiance, though they can form via accretion~\cite{Hui:2019aqm,Clough:2019jpm,Cardoso:2022nzc} or dynamical capture from a DM halo~\cite{Budker:2023sex}. For non-spherical clouds, the technical setup is almost the same as what we outlined above, with the major difference appearing in the non-homogeneous part (source term) of the Klein-Gordon equation. Due to non-sphericity, modes with different angular momentum are coupled, and the source term mode labelled by {$\ell_s, \, m_s$} can induce transitions between cloud states $\{ \ell_i, \, m_i \} \rightarrow \{ \ell_f, \, m_f \}$. Formally, these transitions show up as integrals in the source term of the type
\begin{equation}
    \mathcal{P}_{m_i, \, m_s, \, m_f}^{\ell_i, \, \ell_s, \, \ell_f} = \int d\Omega   Y^*_{\ell_f m_f} Y_{\ell_s m_s} Y_{\ell_im_i} \, ,
\end{equation}
which are only non-zero for particular combinations of $\{ \ell_f, \, \ell_s, \, \ell_i \}$ and $\{ m_f, \, m_s, \, m_i \}$. These define a set of transition rules between states (similar to Clebsch-Gordan coefficients).

Ref.~\cite{Brito:2023pyl} studied in detail equatorial, circular EMRIs immersed in relativistic dipolar clouds ($\ell_i = m_i = 1$), which have the fastest superradiant growth timescale ($\tau^\text{growth}_{\ell m}/M \propto 1/\left(M \mu\right)^{4\ell+ 5} $ and in astrophysical timescales, for the dipolar cloud $\tau^\text{growth} \sim 3 \, \text{years} \, (M/10^5M_\odot)(0.1/M \mu)^9$ ). The qualitative picture is similar to what we obtained for spherical clouds. However, there is a breaking of symmetry whether the EMRI is prograde or retrograde with respect to the cloud's angular momentum (which will be aligned with the central BH spin). Interestingly, for prograde orbits, the EMRI can induce the transition $\{1\, , \, 1\} \rightarrow \{0\, , \, 0\}$, i.e., to a state of lower angular momentum. The net angular momentum transferred from the binary to the cloud can then be negative (see Figs.~3-5 in Ref~\cite{Brito:2023pyl}), making the binary gain angular momentum. If there was no GW emission, this would lead to an outspiral. However, because the gravitational flux dominates over scalar emission, the net effect is just a slower inspiral compared to ``pure'' vacuum.

%

\subsection{LISA Forecasts}
%
\begin{figure}[t]
    \centering \includegraphics[width=0.85\columnwidth]{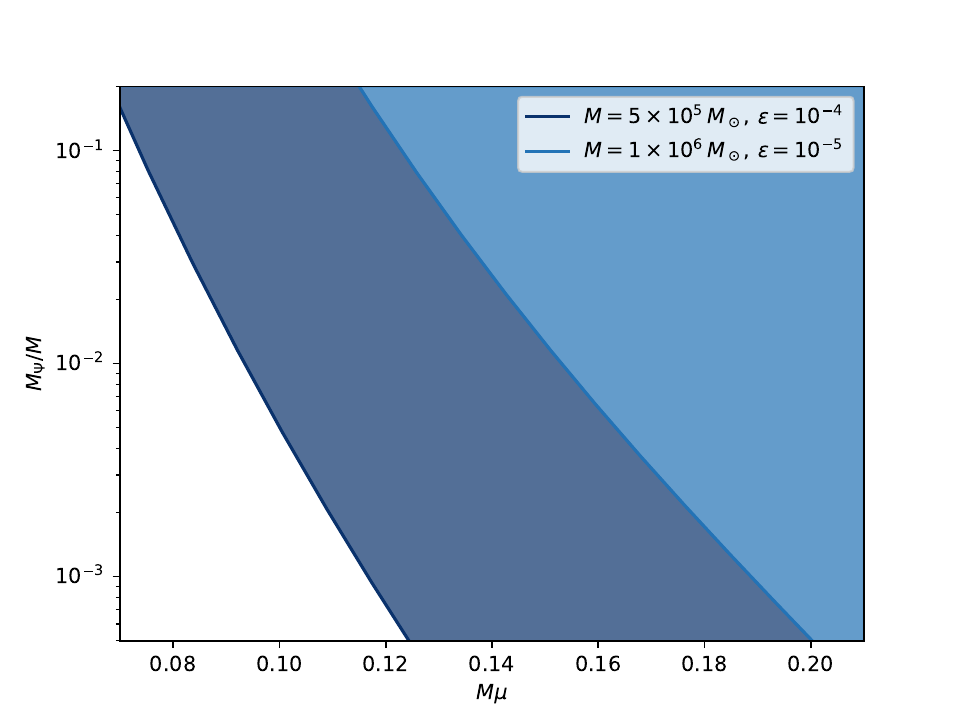}
    \caption{
    Minimum mass of the scalar cloud, $M_\Psi$, which for a given mass of the scalar, $\mu$, leads to a 1 cycle of orbital dephasing after 4 years of observation prior to plunge, for two different EMRIs in prograde orbits immersed in a $\ell_i=m_i=1$ cloud (non-spherical, adapted by Hassan Khalvati from Ref.~\cite{Khalvati:2024tzz}, using results from Ref.~\cite{Brito:2023pyl}) 
    }
    \label{fig:DephasingCloud}
\end{figure}

What are then the consequences of our results to GW astronomy, in particular to LISA? As we just discussed, the interaction of the binary with the environment makes it lose energy at a different rate, which changes the binary's trajectory. This has a direct impact on the phase of the GW signal emitted by the EMRI
\begin{equation}
\phi_\text{GW} = 2 \int_{0}^{T_\text{obs}} \Omega_p(t)dt =  2\int_{0}^{T_\text{obs}} \left( \frac{d\varphi} {dt} \right) dt  =  \int_{0}^{T_\text{obs}} \left(\frac{d\varphi}{dr}\right)\left(\frac{dr}{dE}\right)\left(\frac{dE}{dt}\right)dt \, , 
\end{equation}
which we can measure very accurately. The minimum measurable dephasing depends on different aspects, namely the signal-to-noise ratio (how ``loud'' the GW signal compared to the noise in the detector), but a 1 cycle dephasing is a rule-of-thumb used for detectability (though not a sufficient criteria, see e.g.~\cite{Burke:2023lno}).

In the left panel of Fig.~\ref{fig:DephasingCloud} we show, for a given value of the field mass $M \mu$, the minimum mass of the bosonic cloud that leads to 1 dephasing cycle after 4 years of observation -- the duration of the LISA mission -- that end in a plunge, for two different EMRIs. These are results from a recent study~\cite{Khalvati:2024tzz}, where we used the flux data for relativistic non-spherical clouds from Ref.~\cite{Brito:2023pyl}. Since superradiant clouds can be as massive as $M_\Psi \sim 0.1M$~\cite{Brito:2015oca}, there is an astrophysically relevant region of the parameter space that could lead to measurable effects. LISA could probe clouds with $M\mu \gtrsim 0.07$, which for $M\sim 10^6 - 10^5 \, M_\odot$ corresponds to $\mu \sim 10^{-17}-10^{-16} \, \text{eV}$.  For the system with larger mass ratio ($\varepsilon = 10^{-4}$), the minimum mass is smaller because the initial separation is larger, $r_p^i \sim 20M$ \textit{vs} $r_p^i \sim 10M$ for the system with smaller mass-ratio ($\varepsilon = 10^{-5}$). For larger radius, the relative ratio between the scalar and gravitational flux is larger and hence the effect of the cloud is more significant. In these inspirals we do not hit any resonance, for which the orbital backreaction becomes more involved, since the adiabaticity assumption breaks down~\cite{Tomaselli:2024bdd}. If you want to see more detailed analysis of the detectability of superradiant clouds with EMRIs check Refs.~\cite{Cole:2022yzw, Khalvati:2024tzz}. The upshot is that the prospects for probing ultralight fields with LISA are very exciting! 

\section{Discussion}

In these notes, we focused on the simplest system possible for the theory considered: a circular EMRI around a non-rotating BH, surrounded by a spherical scalar cloud. We end by discussing how to extend this to more complicated setups and what are interesting directions to pursue. 

\noindent{\textbf{\textit{Rotating configurations:}} rotation of the central BH is a necessary ingredient for superradiance. For a rotating background vacuum spacetime, it is still possible to write a master wave equation for the radiative degrees freedom. This is known as the Teukolsky equation~\cite{1973ApJ...185..635T}. However, the metric reconstruction necessary to input the source term of the Klein-Gordon equation becomes much more complicated~\cite{Hollands:2024iqp}. Another route is to consider a slowly-rotating approximation for the background, in which the structure of the equations is more similar to what we have consider~\cite{Pani:2013wsa}. 

\noindent{\textbf{\textit{Fuzzy DM Solitions:}} in the Introduction it was mentioned ultralight fields can also form self-gravitating structures -- boson stars or fuzzy solitons -- which may describe the inner core of DM halos. Even if parasited by a BH in their centre, these objects can be stable over timescales larger than the age of the Universe~\cite{Cardoso:2022nzc}. It is thus natural to consider an EMRI immersed in a fuzzy DM soliton, a problem which has been studied in Ref.~\cite{Duque:2023cac}. The main technical hurdle in this is precisely that the self-gravity of the scalar configuration is non-negligible, so the background spacetime cannot be approximated by only vacuum. To complicate things further, the gravitational sector does not decouple from the scalar one, which means scalar perturbations can also source GWs. In other words, one needs to solve the coupled system of 5 ODEs at the same time. If you want to understand the technicalities behind this better, you can find the \texttt{Mathematica} notebook used in Ref.~\cite{Duque:2023cac}~\href{https://centra.tecnico.ulisboa.pt/network/grit/files/gravitational-waves-boson-clouds/}{here} (\textit{warning}: it is not the most user-friendly/pedagogical tool in place). Note that boson stars are much less compact/dense than boson clouds, so their effect on the inspiral is less significant. Yet, they extend to larger regions and can thus influence the formation of EMRIs (more on this just below). This effect could be studied at the population level instead of searching for deviations of vacuum-GR in a single GW event. 

\noindent{\textbf{\textit{Eccentric/Inclined orbits:}} Throughout our analysis it was assumed the EMRI exists but we have ignored its past history and the impact the EMRI's formation channel has on the configuration of the binary, such as its eccentricity and inclination, and the state of the cloud. The most ``standard'' mechanism for EMRI formation in vacuum is multi-body scattering in dense stellar clusters. In this process, a body is launched to the central SMBH in an almost-plunging, highly eccentric orbit ($e \gtrsim 0.999$), which gradually circularizes via GW emission, and may enter the LISA band still at moderate eccentricity ($e \lesssim 0.7$) and arbitrary inclination~\cite{Babak:2017tow, Broggi:2022udp}. EMRI formation in non-vacuum environments is still poorly understood, but recent results in the context of accretion disks indicate the interaction of the small compact body with the disk assists EMRI formation~\cite{Pan:2021oob, Pan:2021ksp, Derdzinski:2022ltb}, and that these channels could actually dominate the fraction of detectable EMRIs by LISA. What about EMRIs forming immersed in superradiant clouds? Refs.~\cite{Tomaselli:2024bdd, Tomaselli:2024dbw} addressed this problem and found two interesting phenomenological features: first, for most initial orbital configurations, the passage(s) through resonance(s) depletes the cloud almost entirely by the time the compact binary enters the frequency band of GW detectors. The cloud has the highest chance of survival for retrograde orbits (with respect to the BH spin); in any case, resonances leave distinctive features in eccentricity and inclination distribution, with the authors reporting the existence of fixed points in the trajectory evolution of these quantities. This has also been observed in GRMHD simulations of binaries in accretion disks~\cite{Ishibashi:2024wwu, Franchini:2024pgl} and, somewhat mysteriously (at least to me), the value of the fixed point in eccentricity is very similar, $e \sim 0.54$. Considering these results, it appears relevant to extend the studies of EMRIs evolving in relativistic ultralight field environments to generic orbits.

\noindent{\textbf{\textit{Real/Vector fields:}} one can also consider other types of (ultralight) fields. For example, we could have studied \textit{real} scalar fields instead of complex ones. For the former, the energy-momentum tensor is no longer time-independent -- there is no equivalent of the Noether charge in Eq.~\eqref{eq:Noether} -- and the cloud will emit GWs even when isolated. The main technical difference would be then to consider this energy loss in the balance law we wrote for the binary's inspiral. Other possibility is to study vector fields, for which the superradiant instability is stronger ($\tau^\text{growth}_\text{V} \sim 1 \, \text{month} \, (M/10^5M_\odot)(0.1/M \mu)^7$ ) . Here, we would need to substitute the Klein-Gordon equation by the Proca equation for the ultralight vector field $\bm{A}$    
\begin{equation}
    \nabla_\mu \left(  \partial^\mu A^\nu - \partial^\nu A^\mu\right) = \mu^2 A^\nu \, , 
\end{equation}
This problem has been addressed very recently in Ref.~\cite{Cao:2024wby}.  


That is it. With this discussion, I hope to have convinced you that there are multiple open problems in this research programme, where you can work at the intersection of strong-field gravity, particle physics, and astrophysics. 

\section{Acknowledgments}

I want to start by thanking the organizers of the New Horizons for $\Psi$ School and Workshop for the invitation and their extraordinary effort in organizing such a lively meeting, where I had the opportunity to teach but ended up learning much more. I also want to thank Richard Brito, Andrea Maselli, and Rodrigo Vicente whose codes and work I used as a guide to these notes, and to Thomas Spieksma, Giovanni Maria Tomaselli, Yifan Chen, Matteo Della Rocca, Leonardo Gualtieri, Hassan Khalvati, Lorenzo Speri, and Alessandro Santini for fruitful discussions on this subject over the past months. Also, thanks to Thomas and Yifan for comments on an initial version of the manuscript. Finally, a special thanks to Christopher Bauder, Carla Urban, and the rest of the DARK MATTER Berlin team for their inspiring creations, the permission to use pictures of their exhibition on the title page of this manuscript, and their enthusiasm and curiosity in Science and Technology. 

%
\clearpage
\bibliographystyle{unsrt}
\bibliography{main}
\end{document}